\documentclass[journal]{IEEEtran}
\usepackage{graphicx}

\usepackage{cite}

\ifCLASSINFOpdf
\else
\fi
\usepackage{amsmath,amssymb}
\usepackage{siunitx}

\begin{document}
%
\title{Growth and Collapse of an Isolated Bubble Driven by a Single Negative Histotripsy Cycle in Agarose Gel: Stress, Strain, and Strain Rate Fields}
%
%
%

\author{Lauren~Mancia,
        Jonathan~R.~Sukovich,
        Zhen~Xu,~\IEEEmembership{Member,~IEEE,}
        and~Eric~Johnsen
\thanks{L. Mancia and E. Johnsen are with the Department
of Mechanical Engineering, University of Michigan, Ann Arbor,
MI, 48109 USA e-mail: (lamancha@umich.edu; ejohnsen@umich.edu).}
\thanks{J Sukovich and Z Xu are with the Department of Biomedical Engineering, University of Michigan, Ann Arbor, MI, 48109 USA e-mail: (jsukes@umich.edu; zhenx@umich.edu).}
\thanks{}}

\maketitle

\begin{abstract}
Histotripsy relies on cavitation to mechanically homogenize soft tissue. There is strong evidence that the high stresses, strains, and strain rates developed as bubbles grow and collapse contribute to this tissue homogenization. While such stresses and strains have been examined computationally in model systems with assumed constitutive models (e.g., finite-deformation Neo-Hookean model) and viscoelastic properties determined under quasi-static conditions, recent studies proposed that the Quadratic Law Kelvin-Voigt (QLKV) constitutive model, which additionally accounts for  strain stiffening, more accurately represents the viscoelastic response of soft materials subjected to cavitation; this model has also been used to infer viscoelastic properties at high rates. In this work, we use the QLKV model and these properties to calculate the time-dependent stress, strain, and strain rate fields produced during the growth and collapse of individual bubbles subjected to a histotripsy-relevant pressure waveform in agarose gels of 0.3~\% and 1.0~\% concentration and corresponding to actual (past) experiments. We find that, as the gel concentration is increased, strain stiffening manifests in larger elastic stresses and compressive stresses extending into the collapse phase, particularly for the 1.0~\% concentration gel. As a result, the duration of the collapse phase also increases. In comparison with the conventional Neo-Hookean model, the compressive stress has a larger magnitude, extends farther into the surrounding medium, and shows an increased departure from growth/collapse symmetry close to the bubble; all of these effects are magnified in the stiffer gel.

\end{abstract}

\begin{IEEEkeywords}
histotripsy, cavitation, bubble dynamics, ultrasound bioeffects.
\end{IEEEkeywords}

%
\IEEEpeerreviewmaketitle

\section{Introduction}
%
%
%
%
\IEEEPARstart{N}{oninvasive} surgical techniques are broadly favored in the treatment of pathologies affecting nearly all systems of the body as they avoid the numerous complications that may arise during invasive procedures. To this end, hosts of technologies have been investigated and developed to address the wide variety of surgical needs associated with the treatment of different pathologies throughout the body \cite{fry1955ultrasonic, aronow1960use, sweet1960radiofrequency, cooper1962cryogenic, stewart2003focused, sorensen2007radiofrequency, fenner2008analytical, mcdannold2010transcranial, correa2014retrospective}. High-intensity focused ultrasound (HIFU) therapies in particular have garnered significant interest in the noninvasive surgical space owing to their broad applicability, diverse mechanisms of therapeutic action (e.g.\ targeted drug delivery, thermal or mechanical ablation), and the relative safety of ultrasound waves as a delivery mechanism (e.g., compared to ionizing radiation which is dangerous and inherently damaging to tissues \cite{deangelis1989radiation}). Indeed, HIFU therapies, through their various mechanisms of action, have been demonstrated effective in pre-clinical and/or clinical settings for treating pathologies ranging from varicose veins \cite{schultz1989ultrasonic, obermayer2020extracorporeal}, uterine fibroids \cite{stewart2003focused}, and essential tremor \cite{elias2013pilot}, to kidney- and gallstones \cite{chaussy1984extracorporeal, sackmann1988shock}, to stroke \cite{alexandrov2000high} and various cancers \cite{couture2014review}. 


Histotripsy \cite{parsons2006pulsed,xu2005controlled} is a non-thermal HIFU therapy that relies on the targeted generation of cavitation events to mechanically fractionate and destroy soft tissues, and has been demonstrated in a wide variety of tissues \cite{vlaisavljevich2015effects}. Cavitation during histotripsy is generated using short-duration ($\leq 2$ acoustic cycles), high-amplitude ($\gtrsim$\SI{28}{\MPa} \cite{maxwell2013probability}) focused ultrasound pulses delivered from an extracorporeal ultrasound transducer. The high tension produced by focusing megahertz pressure pulses gives rise to clouds of cavitation bubbles, which explosively grow and collapse, thereby destroying the surrounding tissue. There is strong evidence that the high stresses, strains, and strain rates developed during bubble growth and collapse are connected to the tissue homogenization process \cite{vlaisavljevich2016visualizing, mancia2017predicting}. However, although cavitation events during histotripsy can be generated in a controllable and targeted fashion, accurately predicting the bubble cloud dynamics, upon which the induction of damage within the tissue and overall size of the resulting homogenized region depends, remains a challenge and requires an accurate characterization of the pressure field in the focal region and relevant constitutive model for the tissue. The former is challenging due to the presence of many bubbles and nonlinear bubble-bubble and bubble-waveform interactions. The latter is especially difficult at rates relevant to cavitation ($> 10^{5}$ s$^{-1}$), where the constitutive properties of materials are known to exhibit strong rate-dependence; adequately resolving the dynamics of the bubbles necessary to allow these properties to be determined at such high rates remains experimentally challenging. More importantly, acoustically nucleating the types of single spherical bubbles required to compare to predictive models has been difficult to accomplish in practice in a repeatable fashion without the inclusion of external agents to act as seed nuclei \cite{hong2016localized}.

Recent advances in ultrasound transducer technology have enabled the development of experimental setups capable of repeatably growing a single bubble using well defined histotripsy pulses \cite{wilson2019comparative} and data from these experiments were instrumental to validating modeling strategies describing the resulting bubble dynamics in water, including determining the appropriate driving pressure \cite{mancia2020single} and inferring nuclei size distributions \cite{mancia2020measurements}. However, the stress and strain fields in the medium surrounding the bubble, which are readily determined for a Newtonian liquid like water, are not representative of those in soft materials like hydrogels or tissues, which exhibit a viscoelastic response. As such, more sophisticated models, along with well characterized sets of viscoelastic  properties, are required to accurately predict the stresses and strains generated in these materials by cavitation and ultimately inform our understanding of how damage thereby generated. 

By analogy to laser-induced cavitation rheometry  \cite{estrada2018high}, ultrasound-driven cavitation was proposed in \cite{Mancia2020ACR} as a means to characterize the viscoelastic properties of soft materials, including agarose gels, at high rates. That study modeled agarose gel using the  Neo-Hookean model and the higher-order Quadratic Law Kelvin-Voigt (QLKV) model. The former had been favored for representing  large-amplitude bubble growth observed under high-amplitude ultrasound forcing \cite{mancia2019modeling,mancia2017predicting,vlaisavljevich2016visualizing} and deformation of cells in other contexts \cite{peeters2005mechanical}. However, the higher-order QLKV model was shown to achieve superior agreement with experimental data \cite{Mancia2020ACR}. The QLKV model is based on a model originally proposed by Fung \cite{fung2013biomechanics} and includes strain-stiffening effects considered significant at high strain rates developed during cavitation events \cite{estrada2018high,raayai2019capturing}. Notably, the QLKV model inferred shear moduli that were close to their quasi-static measurements whereas shear moduli inferred with the Neo-Hookean model were significantly larger \cite{Mancia2020ACR}. These differences indicate that use of the QLKV model could have significant implications for mechanical damage thresholds (e.g., stresses, strains, and strain rates) considered in prior studies which modeled tissue as a Neo-Hookean material \cite{mancia2017predicting,mancia2019modeling,vlaisavljevich2016visualizing}. 

With a strategy to accurately model single-bubble dynamics in histotripsy \cite{mancia2019modeling} and a constitutive model valid at high rates, the stress and strain fields in surrounding soft matter due to cavitation-bubble growth and collapse can be determined. The present study uses the QLKV model to calculate stress, strain, and strain rate fields produced by the growth and collapse of bubbles driven by histotripsy-relevant waveforms, based on past experiments in 0.3~\% and 1.0~\% agarose gel specimens. Viscoelastic properties of the gels were previously determined using a variant of the Inertial Microcavitation high strain-rate Rheometry (IMR) technique \cite{estrada2018high,yang2020extracting} with the QLKV model \cite{Mancia2020ACR}. Previous studies investigating histotripsy bubble dynamics in viscoelastic media have assumed initial radii are equal to cavitation nucleus sizes in water \cite{mancia2020measurements,mancia2019modeling,maxwell2013probability}; to avoid reliance on this assumption, simulations are initialized with the mean stress-free radius measured for each gel specimen \cite{Mancia2020ACR}. We conclude by considering how the QLKV model impacts prior work demonstrating distinct mechanical origins of maximum compressive stress with increasing distance from the bubble \cite{mancia2017predicting}.

\section{Methods}

In this work, we select two representative data sets from past experiments of ultrasound-generated growth and collapse of a single cavitation bubble in agarose  \cite{wilson2019comparative}, one in a 0.3~\% gel and the other in a 1.0~\% gel. We then simulate the corresponding bubble growth and collapse using a single-bubble model to yield the time history of the bubble radius, from which we calculate the associated radial stress, strain, and strain rate fields. We describe here the experiments, model, and field calculation.

\subsection{Experiments}

In the present work, we select one representative data set from the ensemble of 19 experiments in 0.3~\% agarose and one representative data set from the ensemble of 20 experiments 1.0~\% agarose from the experiments of \cite{wilson2019comparative}. Briefly, those
experiments were carried out in a open-topped, \SI{10}{\cm} diameter spherical histotripsy array comprised of 16 focused acoustic transducer elements with a center frequency of \SI{1}{\MHz}. During experiments the array was filled with deionized water, filtered to \SI{2}{\um} and degassed to \SI{4}{\kPa}. Agarose gel samples measuring \SI{2.5}{\cm} in diameter and \SI{7.5}{\cm} in length, with concentrations (w/v) of $0.3$~\% and $1.0$~\% \cite{vlaisavljevich2015effects}, were prepared for cavitation experiments and inserted into the transducer for nucleation via the opening at the top. For reference, the quasi-static shear moduli of the $0.3$~\% and $1.0$~\% gel samples were \SI{3.4}{\kPa} and \SI{65.1}{\kPa}, respectively. The acoustic pulses responsible for nucleating single spherical bubbles in the gel samples measured 1.5 acoustic cycles (\SI{1.5}{\us}) and contained only a single rarefactional pressure half-cycle with a peak focal pressure of \SI{-24}{\MPa} \cite{maxwell2013probability}. All bubbles were nucleated \SI{\geq 5}{\mm} from the edge of the gel samples to avoid potential influences from boundary effects on the resulting bubble dynamics. The dynamics of the nucleated bubbles were monitored from their inception until the time of first collapse using a high speed camera in combination with an adaptive, multi-flash-per-camera-exposure illumination technique \cite{sukovich2020cost}. Each of the two selected data sets is the realization closest to the mean of each ensemble. The time history of the bubble radius is shown in Fig.~\ref{fig:rvt} with circle markers corresponding to the $0.3$ \% data set and square markers corresponding to the $1.0$ \% data set, indicating excellent agreement between the model results and the experiments. The emphasis of the present study lies in this first growth and collapse. A full description of the experiments can be found in \cite{wilson2019comparative}.

\subsection{Theoretical Model and Numerical Methods}

\begin{figure*}
\begin{center}
  \includegraphics[width=\textwidth]{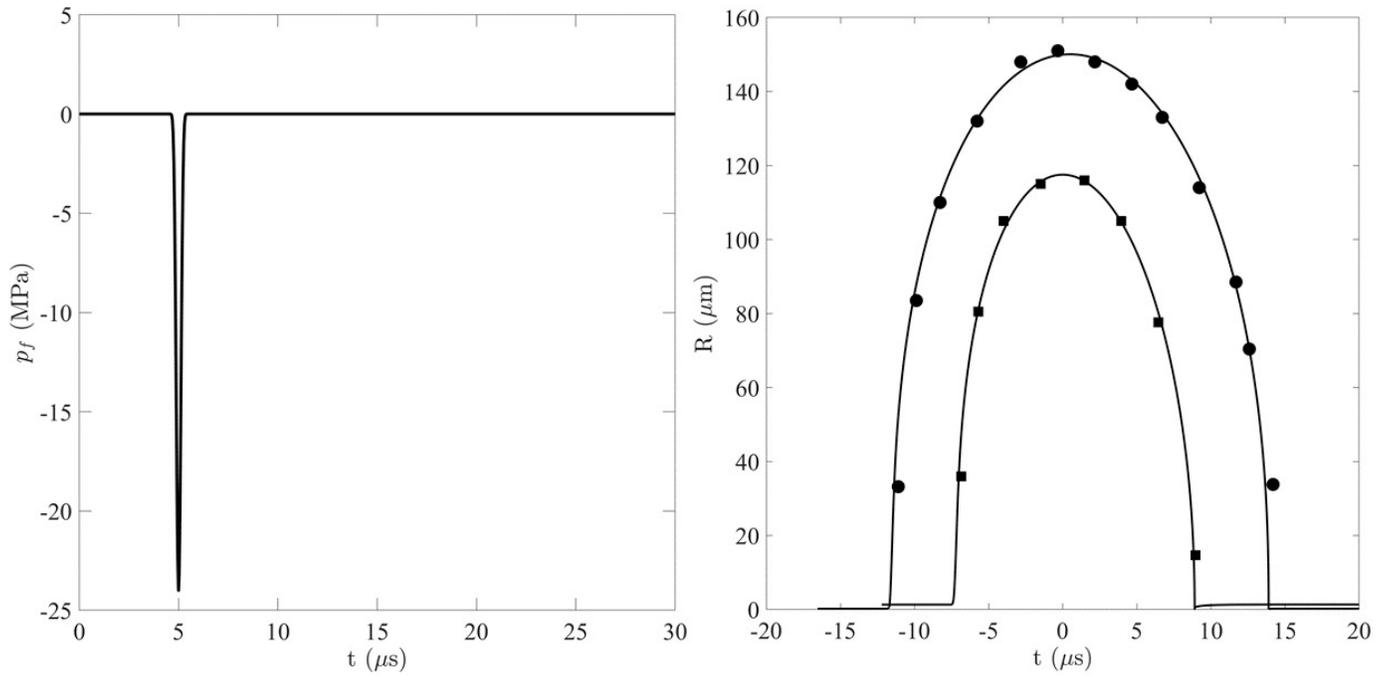}
  \caption{Left: Validated analytic waveform used in simulations. Right: Representative radius vs. time data sets for 0.3~\% (circle markers) and 1.0~\% (square markers) gels time-shifted such that maximum radius occurs at $t = 0$. Traces correspond to simulation results obtained with inferred gel parameters in Table~\ref{table:Props}.}
  \label{fig:rvt}
  \end{center}
\end{figure*}

Following past studies \cite{mancia2019modeling,bader2018influence,mancia2017predicting,vlaisavljevich2016effects,vlaisavljevich2016visualizing,vlaisavljevich2015effects}, our modeling approach considers the dynamics of a single spherical bubble in an infinite, homogeneous viscoelastic medium. The time-history of the bubble radius $R(t)$ is governed by the Keller-Miksis equation \cite{keller1980bubble}:
\small
\begin{align}
\label{KM}
\begin{split}
&\left(1 -\frac{\dot{R}}{c_{\infty}}\right)R\ddot{R}+\frac{3}{2}
\left(1 - \frac{\dot{R}}{3c_{\infty}}\right)\dot{R}^2 = \\
&\frac{1}{\rho_{\infty}}\left(1 +\frac{\dot{R}}{c_{\infty}}
+\frac{R}{c_{\infty}}\frac{d}{dt}\right)\Biggl[p_b-p_{\infty}\Biggl(t + \frac{R}{c_{\infty}}\Biggr)-\frac{2S}{R} + J \Biggr],
\end{split}
\end{align}
\normalsize
where the sound speed, $c_{\infty}$, density, $\rho_{\infty}$, and surface tension, $S$, are fixed at the values given in prior work \cite{wilson2019comparative}, and $J$ is the  integral of the deviatoric contribution of the stresses in the surroundings. The far-field driving pressure, $p_{\infty}(t)$, is a sum of the constant ambient pressure, $P_{0}$, and an analytic function representative of a histotripsy pulse \cite{mancia2019modeling,mancia2017predicting,mancia2020single}.  This pressure waveform has an amplitude of $-24$ MPa, as shown in Fig.~\ref{fig:rvt}. The bubble is homobaric with pressure $p_b(t)$; the calculation of the bubble pressure is coupled to the energy balance partial differential equation, which is discretized inside the bubble to more accurately account for energy tranport \cite{prosperetti1991thermal,prosperetti1988nonlinear,kamath1993theoretical}.  The bubble-gel interface is assumed to be impervious to gas, and gel surrounding the bubble remains at a constant ambient temperature of $25$ $^{\circ}$C. These assumptions have been adopted by previous authors  \cite{prosperetti1991thermal,prosperetti1988nonlinear,kamath1993theoretical,warnez2015numerical,mancia2020single, mancia2019modeling} and are acceptable for modeling the single cycle of growth and collapse typically resolved in histotripsy-relevant cavitation experiments \cite{wilson2019comparative,mancia2020single}. As in \cite{Mancia2020ACR}, the Quadratic Law Kelvin-Voigt (QLKV) constitutive relation \cite{yang2020extracting} is used to relate stresses and strains in agarose gels. In this model, the stress integral $J$ takes the following form:
\small
\begin{align}
\begin{split}
\label{eq:SS}
J &= -\frac{4\mu \dot{R}}{R} + \frac{G(3\alpha - 1)}{2}\left[ 5 - 4\left(\frac{R_0}{R}\right) - \left(\frac{R_0}{R}
  \right)^4\right]\\
& + 2G\alpha\left[ \frac{27}{40} + \frac{1}{8}\left(\frac{R_0}{R}\right)^8 +\frac{1}{5}\left(\frac{R_0}{R} \right)^5
+\left(\frac{R_0}{R}\right)^2 -\frac{2R}{R_0}\right],
\end{split}
\end{align}
\normalsize
where $R_0$ is the stress-free radius corresponding to a reference configuration; departures from this radius give rise to restoring elastic stresses. The viscoelastic properties of the gel specimens include  viscosity, $\mu$, shear modulus, $G$, and  stiffening parameter, $\alpha$, and are taken to be constant over the course of the simulation. When $\alpha = 0$, the QLKV model reduces to the Neo-Hookean model \cite{gaudron2015bubble}. The values of these properties obtained by Mancia et al. \cite{Mancia2020ACR} for the Neo-Hookean and QLKV models are given in Table \ref{table:Props}.

The discretized equations are solved numerically as described previously \cite{mancia2019modeling} with the MATLAB \textit{ode15s}  time-marching scheme \cite{shampine1997matlab,shampine1999solving} and second-order central differences for spatial derivatives in the energy equation \cite{estrada2018high,barajas2017effects}. The $R(t)$ solutions obtained for the two representative data sets considered in this study are shown as the line traces in Fig. \ref{fig:rvt}. Stresses, strains, and strain rates are calculated using the $R(t)$ results obtained for each gel concentration as described in the following section (Sect. \ref{damagemech}).

\begin{table}[t!]
\small
  \caption{\ Properties inferred from representative 0.3~\% and 1~\% agarose experiments \cite{Mancia2020ACR} shown in Fig.~\ref{fig:rvt}.}
  \label{table:Props}
  \begin{tabular*}{0.50\textwidth}{@{\extracolsep{\fill}}lllll}
    \hline
    Model & $G$ (kPa) & $\alpha$ ($10^{-2}$) & $\mu$ (Pa$\cdot$s) & $R_0$ ($\mu$m) \\
    \hline\hline
    \ 0.3\% gel \\
    \hline
    NH  & 9.1  & 0 & 0.077 & 0.25 \\
    QLKV  & 0.44 &  1.5 & 0.079  & 0.21 \\
    \hline\hline
    \ 1\% gel \\
    \hline
    NH  & 31 & 0 & 0.15  & 1.3 \\
    QLKV  & 7.5  & 2.8  & 0.15  &  1.3  \\
    \hline
  \end{tabular*}
\end{table}

\subsection{Calculation of Stresses, Strains, and Strain rates}
\label{damagemech}

Stress, strain, and strain rate fields are calculated as described previously \cite{mancia2019modeling} using the QLKV and Neo-Hookean models. For all field quantities, the original, $r_0$ and current, $r$, radial coordinates are related by:
\begin{equation}\label{eq:r_r0}
  r_0(r,t)=\sqrt[3]{r^{3}-R^{3}+R_0^{3}},
\end{equation}
where $R_0$ is the stress-free radius. We consider only the radial components of total deviatoric stress, $\tau_{rr}$, which has a simple relation to hoop stresses, $\tau_{rr} = -2\tau_{\theta\theta}$ in incompressible gels. As in the Neo-Hookean model \cite{mancia2017predicting}, total stress in the QLKV model can be expressed as a sum of its elastic, $\tau_{rr}^E$, and viscous, $\tau_{rr}^V$, components:
\begin{equation}\label{eq:totstress}
  \tau_{rr}=\tau_{rr}^E + \tau_{rr}^V,
\end{equation}

\small
\begin{align}\label{eq:estress}
\tau_{rr}^E= \frac{2G}{3}\left[1 + \alpha\left(\left(\frac{r_0}{r}\right)^4+2\left(\frac{r}{r_0}\right)^2-3\right)\right]\left[\left(\frac{r_0}{r}\right)^4-\left(\frac{r}{r_0}\right)^2\right] 
\end{align}
\normalsize

\begin{align}\label{eq:vstress}
\tau_{rr}^V=-4\mu\frac{R^2\dot{R}}{r^3}.
\end{align}

\noindent
Strain fields are calculated using the Hencky (true strain) definition:
\begin{align}\label{eq:strain}
E_{rr} &= -2\ln\left(\frac{r}{r_0}\right).
\end{align}

\noindent
Strain rate fields are calculated using a time derivative of Eq.~\ref{eq:strain}:
\begin{align}\label{eq:strainrate}
\dot{E}_{rr} &= -2\frac{R^2\dot{R}}{r^3}.
\end{align}

\begin{figure*}
\begin{center}
  \includegraphics[width=\textwidth]{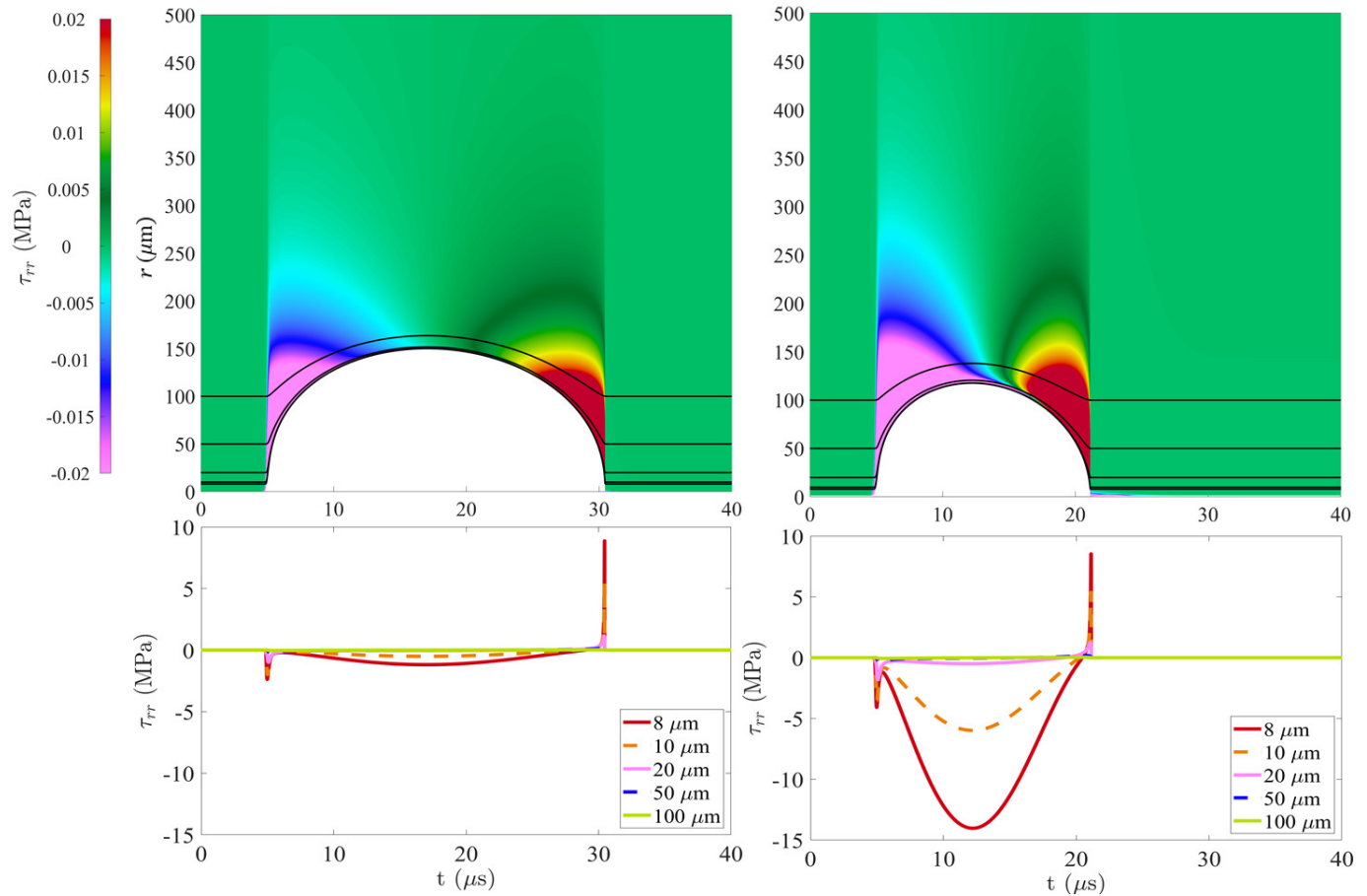}
  \caption{Total deviatoric radial stress fields developed in the 0.3~\% (left) and 1.0~\% (right) gels. White region corresponds to tissue deformation with respect to distance from the stress free radius as a function of time. Black lines overlaying the color plots correspond to Lagrangian paths taken by particles starting 8, 10, 20, 50, and and 100 $\mu$m from the bubble center. Plots in the lower row show the  magnitude of total stress along these paths.}
  \label{fig:TotStress}
  \end{center}
\end{figure*}

\section{Results}

\subsection{Stress fields}

The radial dynamics and total deviatoric radial stress fields obtained with Eq.~\ref{eq:totstress} and corresponding to the bubble growth shown in Fig.~\ref{fig:rvt} are shown in Fig.~\ref{fig:TotStress} for each gel concentration. The white region represents the bubble, which displaces the gels thereby causing their deformation and the associated stresses. The bubble achieves a larger maximum radius in the less stiff 0.3~\% gel, which also gives rise to a longer collapse time.  Stresses are initially compressive (negative) during bubble growth and become tensile (positive) as the bubble collapses to its minimum radius. The stresses are largest near the bubble wall and decrease with distance from the bubble. In the stiffer 1.0~\% gel, the compressive stress has larger magnitude, extends farther into the surrounding medium, and shows an increased departure from growth/collapse symmetry as evidenced by the relatively large tension near the bubble after reaching maximum radius. Lagrangian paths initially located at 8, 10, 20, 50, and 100 microns from the bubble center are indicated by the solid black line overlay in the contours. Total stress experienced along each Lagrangian path is shown below the corresponding contour plot. As the bubble expands, the separation distance between these different trajectories becomes smaller as a material element becomes smaller in the radial direction and elongates in the hoop direction. Local maxima in stress occur at the onset of bubble growth, at bubble collapse (minimum radius), and at the point of maximum bubble radius. At 10 microns from the bubble center, the absolute maximum stress is tensile and occurs at collapse (minimum radius) in the $0.3$~\% gel; in contrast, the absolute maximum stress is compressive and occurs at maximum bubble radius in the 1.0~\% gel.

\begin{figure*}[t]
\begin{center}
  \includegraphics[width=\textwidth]{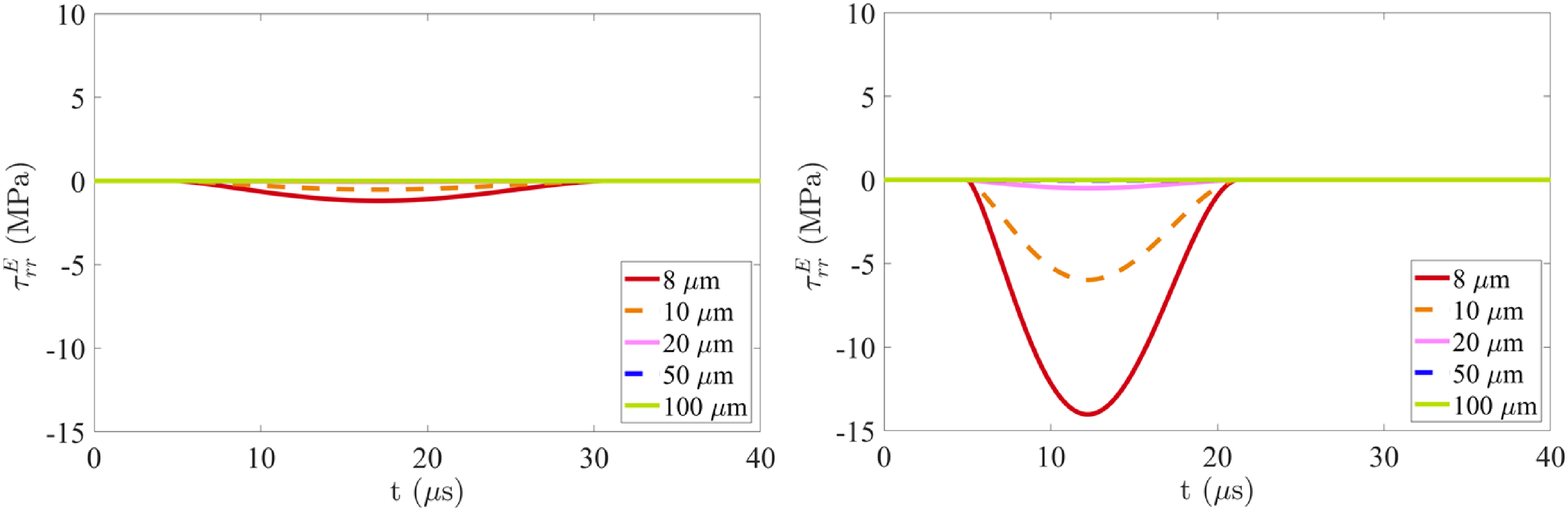}
  \caption{Elastic components of deviatoric radial stress as a function of time experienced by particles starting at various distances from the bubble center for 0.3~\% (left) and 1.0~\% (right) gels.}
  \label{fig:EStress}
  \end{center}
\end{figure*}

\begin{figure*}
\begin{center}
  \includegraphics[width=\textwidth]{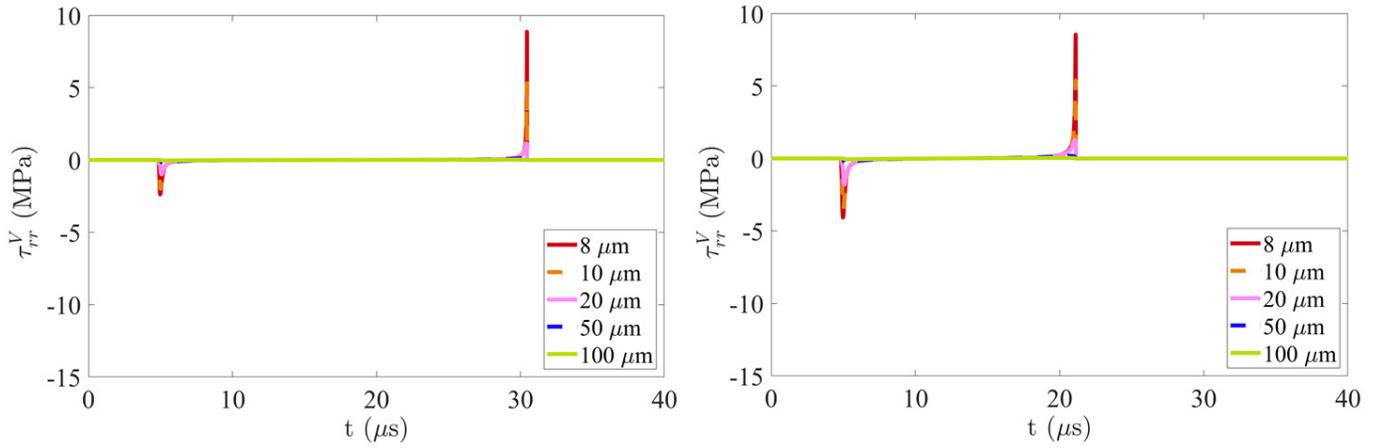}
  \caption{Viscous components of deviatoric radial stress as a function of time experienced by particles starting at various distances from the bubble center for 0.3~\% (left) and 1.0~\% (right) agarose gels.}
  \label{fig:VStress}
  \end{center}
\end{figure*}

To illustrate the relative contribution of viscous and elastic stresses, Figs.~\ref{fig:EStress} and \ref{fig:VStress} show the elastic and viscous components of the total stress calculated using Eqs.~\ref{eq:estress} and \ref{eq:vstress} along each Lagrangian path depicted in Fig.~\ref{fig:TotStress}.  Elastic stress is largest at maximum bubble radius while peaks in viscous stress occur at the onset of bubble growth and at bubble collapse to minimum radius. Significantly larger elastic stresses are developed in the stiffer 1.0~\% gel while viscous stress is maximized in the $0.3$~\% gel. By comparing the elastic and viscous stresses  to the total stress plots in Fig. \ref{fig:TotStress}, it is clear that the absolute maximum compressive stress occurs at maximum bubble radius and is elastic in origin in the 1.0~\% gel, while the absolute maximum compressive stress occurs at the onset of bubble growth in the 0.3~\% gel and is viscous in origin.

\subsection{Strain \& Strain Rate Fields}

\begin{figure*}[t]
\begin{center}
  \includegraphics[width=\textwidth]{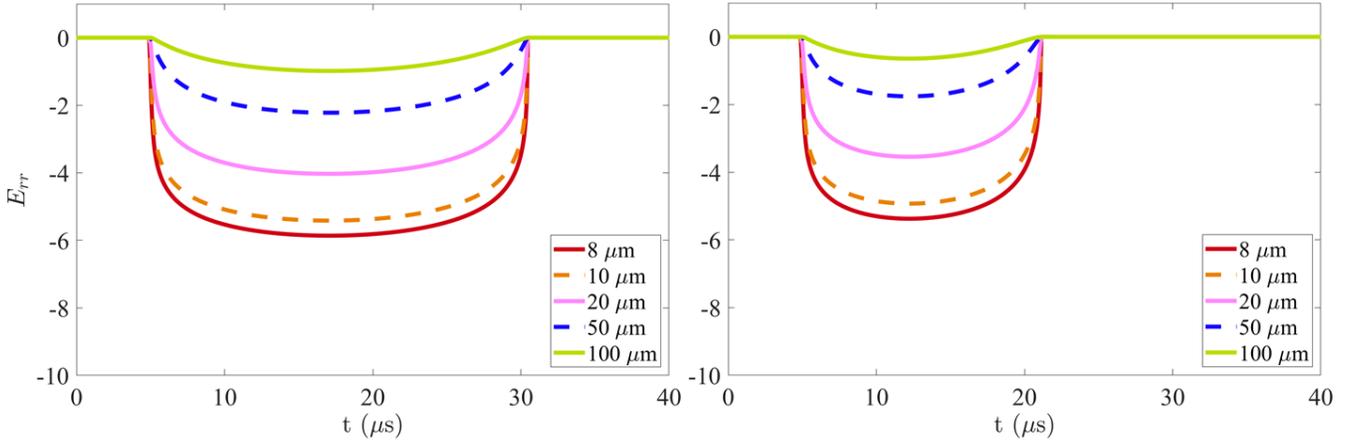}
  \caption{Strain as a function of time experienced by particles starting at various distances from the bubble center for 0.3~\% (left) and 1.0~\% (right) agarose gels.}
  \label{fig:Strain}
  \end{center}
\end{figure*}

\begin{figure*}
\begin{center}
  \includegraphics[width=\textwidth]{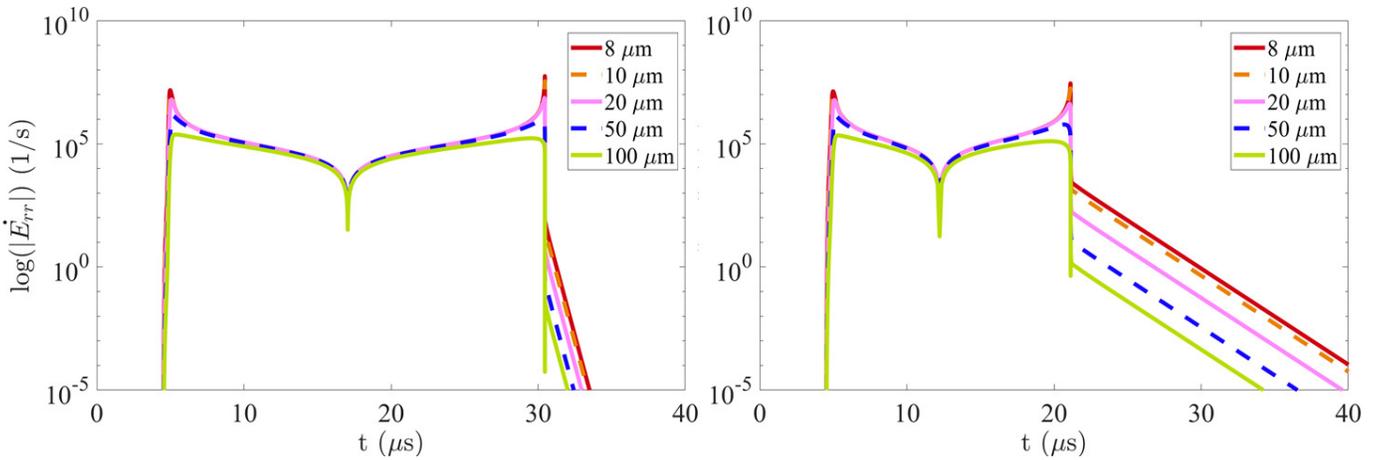}
  \caption{Strain rate as a function of time experienced by particles starting at various  distances from the bubble center for 0.3~\% (left) and 1.0~\% (right) agarose gels.}
  \label{fig:StrainRate}
  \end{center}
\end{figure*}

The radial strains and strain rates experienced in each gel along Lagrangian paths initially located at 8, 10, 20, 50, and 100 microns from the bubble center are calculated using Eqs.~\ref{eq:strain} and \ref{eq:strainrate} and are shown in Figs.~\ref{fig:Strain} and \ref{fig:StrainRate}, respectively. Strains are purely compressive 5 microns from the bubble center and are largest at maximum bubble radius. A slightly larger maximum strain is achieved in the $0.3$~\% gel, reflecting the larger maximum bubble radius relative to the equilibrium radius in this case. The strain rate magnitude is largest at the onset of bubble growth and at collapse to minimum bubble radius. Strain rates then follow a power law decay in space. Although comparable strain rates are reached in both gels, a slightly larger maximum strain rate at collapse is observed in the $0.3$~\% gel. 

\section{Discussion}

The investigation of cavitation damage mechanisms in tissue and tissue-like media has been limited by uncertainties in viscoelastic properties. Until our work integrating single-bubble dynamics modeling and cavitation experiments to determine viscoelastic properties at high rates \cite{Mancia2020ACR}, previous studies \cite{mancia2019modeling,mancia2017predicting,bader2018influence} had to rely on values originating from quasi-static measurements. An additional modeling uncertainty pertaining to the composition of the material is the initial radius size--or the nucleus/nidus size, as evidenced by the wide range of values assumed for histotripsy bubbles in different media \cite{bader2019whom,edsall2020bubble,mancia2017predicting}. Our choice to initialize our simulations using the stress-free radius is consistent with past observations that the gel is likely to fracture due to the large stretch during explosive bubble growth \cite{mancia2019modeling}; we note that the calculated stresses and strains are dependent upon this quantity. Though it is not possible to infer the actual nucleus size, our approach does not require modeling of this rupturing process \cite{movahed2016cavitation}.

\begin{figure*}
\begin{center}
  \includegraphics[width=\textwidth]{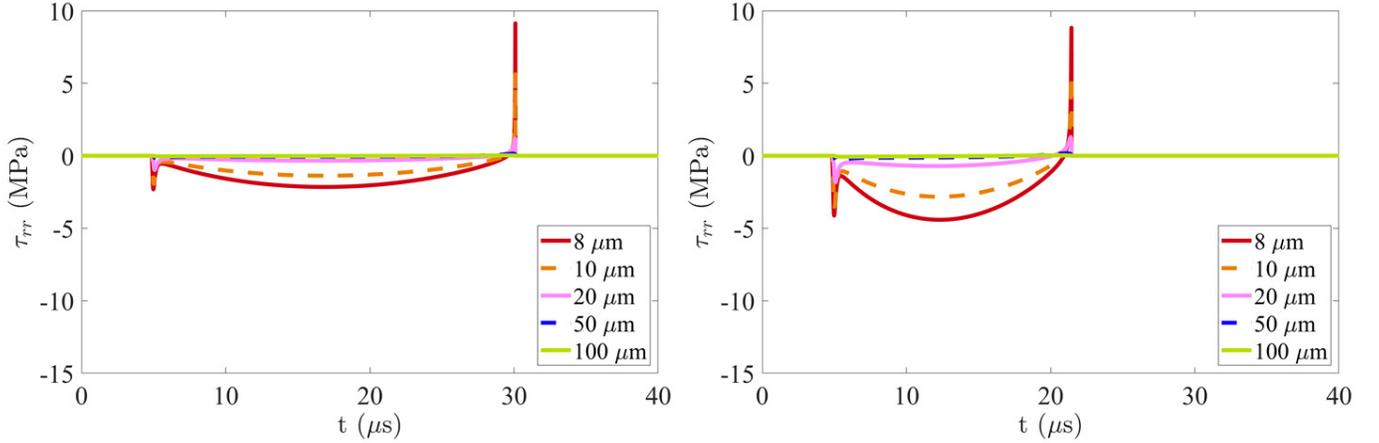}
  \caption{Total deviatoric radial stress as a function of time experienced by particles starting at various  distances from the bubble center for 0.3~\% (left) and 1.0~\% (right) agarose gels obtained with a Neo-Hookean model. These plots are in contrast to the QLKV model stress plots in Fig.~\ref{fig:TotStress}.}
  \label{fig:NHstress}
  \end{center}
\end{figure*}

The distinguishing feature of the QLKV model is strain stiffening--increased stiffness at large strains. To better appreciate this effect, Fig.~\ref{fig:NHstress} shows the time history of the total deviatoric stress when computing the stresses using the Neo-Hookean model, which does not account for strain stiffening; this figure should be compared to the stress traces in Fig.~\ref{fig:TotStress}. In the less stiff, 0.3~\% gel, the stresses are comparable; the Neo-Hookean compressive (elastic) stresses at maximum radius calculated with the Neo-Hookean model are slightly larger than those computed with the QLKV model. However, for the stiffer 1.0~\% gel, the QLKV yields significantly larger stresses. This result highlights the importance of strain stiffening at large deformations in stiffer materials. 

A consequence of this strain stiffening is a more pronounced asymmetry of the bubble radius in time between growth to maximum radius and maximum radius to collapse as the agarose concentration is increased. Experiments indicate that the collapse phase is longer than the growth phase when scaled by the maximum radius and Rayleigh collapse time \cite{wilson2019comparative}. When examining the computed stress fields, it is clear that the region of high compressive stress (e.g., the pink region in Fig.~\ref{fig:TotStress}) extends beyond the maximum radius. By Newton's third law, this implies that there is resistance to the collapsing bubble due to the elastic stresses, which is expected to lead to a longer collapse. 

As noted in previous studies based on a Neo-Hookean model \cite{mancia2017predicting,mancia2019modeling}, stress maxima may be of elastic or viscous origin. Elastic stresses are largest at maximum bubble radius, when the strain is largest. Viscous stresses are largest at the onset of bubble growth and at bubble collapse, when the strain rate is largest. Although maximum tensile stress is always of viscous origin, maximum compressive stress can be of viscous or elastic origin \cite{mancia2017predicting}. These same observations hold for the stresses calculated with the QLKV model. The compressive stresses at 8 and 10 microns from the bubble center in Figs.~\ref{fig:TotStress}, \ref{fig:EStress}, and \ref{fig:VStress} reflect these different regimes of mechanical behavior. In the $0.3$~\% gel, the maximum compressive stress occurs at the onset of bubble growth and is viscous origin. At the same distances in the 1.0~\% gel, the maximum compressive stress occurs at maximum bubble radius and is elastic in origin. 

\begin{figure}
  \includegraphics[width=0.5\textwidth]{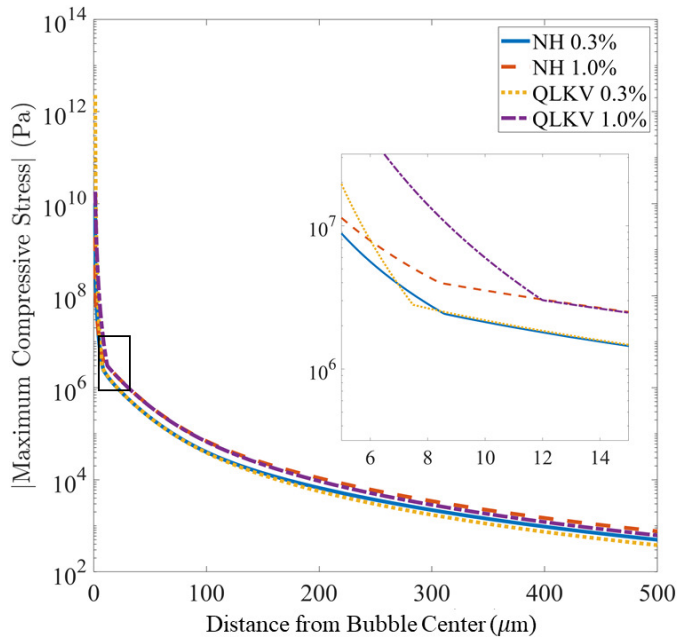}
  \caption{Maximum compressive stress as a function of distance from the bubble center (starting at the stress-free radius) in the $0.3$~\% and $1.0$~\% gels calculated using the Neo-Hookean (NH) and QLKV models. Black box outlines elastic-to-viscous transition points (abrupt changes in slope enlarged in inset). }
  \label{fig:NHvQLKV}
\end{figure}

The origin of maximum compressive stresses in gels modeled with the QLKV model can be further explored by examining the magnitude of maximum compressive stress at increasing distances from the bubble. Fig.~\ref{fig:NHvQLKV} shows the magnitude of maximum compressive stress over the course of the simulation as a function of distance from the stress-free radius for each model and gel specimen. All maximum stress traces exhibit an abrupt change in slope in the boxed region shown enlarged at right. These 'kinks' in the traces correspond to the distance from the bubble center at which compressive stress of viscous origin first exceeds compressive stress of elastic origin. This has been called the elastic-to-viscous transition distance in previous studies of stresses obtained with the Neo-Hookean model  \cite{mancia2017predicting,mancia2019modeling}. The elastic-to-viscous transition distances calculated using each viscoelastic model in each gel specimen are given in Table \ref{table:xev}. The Neo-Hookean model results in nearly equivalent transition distances in each gel concentration because the 1.0~\% gel has a larger viscosity and a larger shear modulus than the $0.3$~\% gel. The transition distance is slightly smaller in the 1.0~\% gel due to its significantly larger shear modulus. In the QLKV model, the elastic-to-viscous transition occurs closer to the stress-free radius for the less stiff $0.3$~\% gel. Additionally, the QLKV model results in a larger elastic-to-viscous transition distance in the 1.0~\% gel. This behavior is consistent with expectations that the higher-order elastic effects of the QLKV model give rise to larger elastic stresses that persist to a greater distance from the bubble.

The elastic-to-viscous transition distances predicted with the QLKV and Nee-Hookean models, while distinct, are separated by less than 4 microns. Beyond these locations, the models display identical stress behavior that reflects their shared viscous stress term and similar strain rates (Eq.~\ref{eq:vstress}). Use of the QLKV model is most likely to affect damage predictions in stiffer gels, in which the higher-order elastic effects encompassed by the $\alpha$-dependent terms of Eq.~\ref{eq:totstress} contribute to markedly larger stresses prior to the elastic-to-viscous transition. This behavior is evident when the total deviatoric radial stress fields obtained with the QLKV model (Fig.~\ref{fig:TotStress}) are compared to those obtained with the Neo-Hookean model (Fig.~\ref{fig:NHstress}). Also, the $0.3$~\% gel stress trace at $8$ microns from the bubble center reveals a maximum compressive elastic stress at maximum bubble radius that is comparable to the maximum compressive viscous stress. This behavior reflects the slightly farther transition distance obtained with the Neo-Hookean model in this case. In contrast, Neo-Hookean stress traces in the stiffer $1$~\% gel have significantly smaller maximum elastic stresses than those obtained with the QLKV model. The stiffer gel plot also demonstrates the comparable magnitudes of elastic and viscous compressive stresses at $8$ microns, reflecting a transition distance closer to the bubble center in the Neo-Hookean case.

\begin{table}
\small
  \caption{ Elastic-to-viscous transition distance in microns obtained with each model in each gel specimen. }
  \label{table:xev}
  \begin{tabular*}{0.50\textwidth}{@{\extracolsep{\fill}}lll}
    \hline
    Model & 0.3\% gel & 1\% gel\\
    \hline\hline
    NH  & 8.54  & 8.49 \\
    QLKV  & 7.49 &  12.0 \\
  \end{tabular*}
\end{table}

\section{Conclusions}

Recent studies proposed that the Quadratic Law Kelvin-Voigt (QLKV) constitutive model, which accounts for strain stiffening, more accurately represents the viscoelastic response of soft materials subjected to cavitation than previously used models (e.g., finite-deformation Neo-Hookean model); this model has also been used to measure viscoelastic properties at high rates. In this work, we use the QLKV model and these properties to calculate the time-dependent stress, strain, and strain rate fields produced during the growth and collapse of individual bubbles subjected to a histotripsy-relevant pressure waveform in agarose gels of 0.3~\% and 1.0~\% concentration and corresponding to actual (past) experiments. We find that, as the gel concentration is increased, strain stiffening manifests in larger elastic stresses and compressive stresses extending into the collapse phase, particularly for the 1.0~\% concentration gel. As a result, the duration of the collapse phase also increases. In comparison with the conventional Neo-Hookean model, the compressive stress has a larger magnitude, extends farther into the surrounding medium, and shows an increased departure from growth/collapse symmetry close to the bubble; all of these effects are magnified in the stiffer gel. In the future, more detailed experimental observations of the rupture of soft materials during explosive bubble growth would greatly benefit the development of models for cavitation damage to soft matter.


%



\section*{Acknowledgment}

This  work  was  supported  by  ONR  Grant  No.\  N00014-18-1-2625  (under Dr. Timothy Bentley).

\ifCLASSOPTIONcaptionsoff
  \newpage
\fi



\bibliographystyle{IEEEtran}
\bibliography{./HSEBib}
%



%

\begin{IEEEbiography} [{\includegraphics[width=1in,height=1.25in,clip,keepaspectratio]{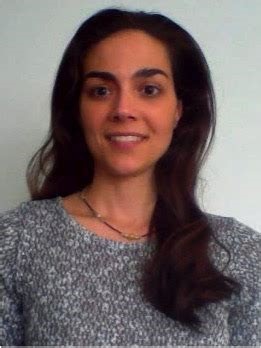}}]{Lauren Mancia}
received the B.S.E. degree in engineering physics from the University of Michigan, Ann Arbor, MI, USA, in 2012.  She completed post-baccalaureate studies in the biological sciences at Wayne State University, Detroit, MI, USA in 2013. She returned to the University of Michigan to complete the M.S.E. degree in mechanical engineering in 2015.  Her M.S.E. studies were funded through a National Science Foundation Graduate Research Fellowship.  She then began medical school at the University of Michigan and joined the Medical Scientist Training Program in 2017. In 2020, she successfully defended her Ph.D. thesis in mechanical engineering and will complete her M.D. degree in 2021.   Her research interests include high strain-rate injury mechanics and focused ultrasound therapies.
\end{IEEEbiography}

\begin{IEEEbiography}[{\includegraphics[width=1in,height=1.25in,clip,keepaspectratio]{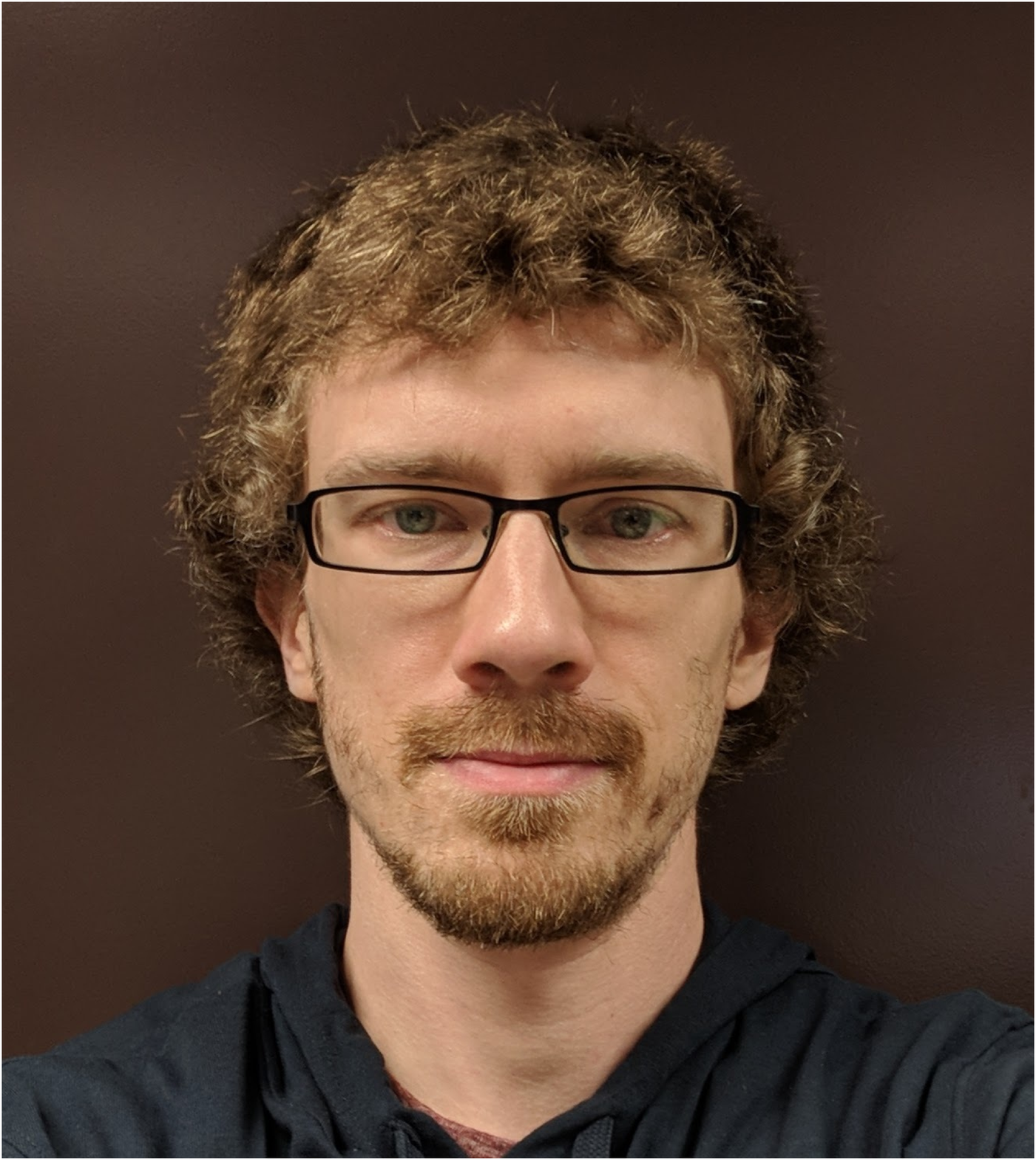}}]{Jonathan R. Sukovich} received the B.S. and Ph.D. degrees in mechanical engineering from Boston University, Boston, MA, USA, in 2008 and 2013, respectively, where he studied laser interactions with water at high pressures and phenomena associated with high-energy bubble collapse events. He joined the University of Michigan, Ann Arbor, MI, USA, in the summer of 2013 to study histotripsy for brain applications. He is currently an Assistant Research Scientist with the Department of Biomedical Engineering, University of Michigan. His research interests include high-energy bubble collapse phenomena, focused ultrasound therapies, and acoustic cavitation.
\end{IEEEbiography}

\begin{IEEEbiography}[{\includegraphics[width=1in,height=1.25in,clip,keepaspectratio]{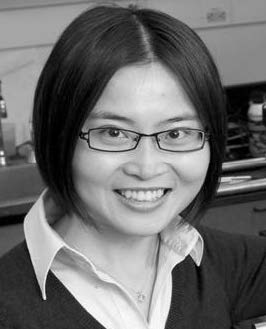}}]{Zhen Xu}
(Member, IEEE) received the B.S.E.degree (Hons.) in biomedical engineering from Southeast University, Nanjing, China, in 2001, and the M.S. and Ph.D. degrees in biomedical engineering from the University of Michigan, Ann Arbor, MI, USA, in 2003 and 2005, respectively. She is currently an Associate Professor with the Department of Biomedical Engineering, University of Michigan. Her research is focused on ultrasound therapy, particularly the applications of histotripsy for noninvasive surgeries. Dr. Xu received the IEEE Ultrasonics, Ferroelectrics, and Frequency Control Society Outstanding Paper Award in 2006, the American Heart Association Outstanding research in Pediatric Cardiology in 2010, the National Institutes of Health New Investigator Award at the First National Institute of Biomedical Imaging and Bioengineering Edward C. Nagy New Investigator Symposium in 2011, and the Frederic Lizzi Early Career Award from the International Society of Therapeutic Ultrasound in 2015. She is also an Associate Editor of the IEEE TRANSACTIONS ON ULTRASONICS, FERROELECTRICS, AND FREQUENCY CONTROL (UFFC).
\end{IEEEbiography}

\begin{IEEEbiography}[{\includegraphics[width=1in,height=1.25in,clip,keepaspectratio]{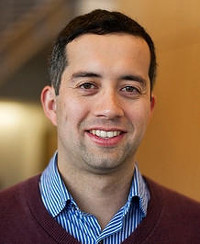}}]{Eric Johnsen}
received the B.S. degree from the University of California at Santa Barbara, Santa Barbara, CA, USA, and the M.S. and Ph.D. degrees from the California Institute of Technology, Pasadena, CA, USA, all in mechanical engineering. He was a Postdoctoral Fellow at the Center for Turbulence Research, Stanford University, Stanford, CA, USA. He is currently an Associate Professor in the Mechanical Engineering Department, University of Michigan, Ann Arbor, MI, USA. His group's research draws from applied mathematics, numerical/physical modeling and high-performance computing to develop numerical simulations and modeling techniques to investigate the basic physics underlying complex multiscale and multiphysics flows, with a focus on multiphase flows, turbulence, shocks, and high-energy-density physics. His work finds applications in biomedical engineering (diagnostic and therapeutic ultrasound, cardiovascular flow, traumatic brain injury), transportation engineering (aeronautical, automotive, naval, hypersonics), astrophysics, and the energy sciences (inertial fusion, nuclear energy). Dr. Johnsen received  the National Science Foundation CAREER Award and the Office of Naval Research Young Investigator
Award, and is an associate fellow of the AIAA.\end{IEEEbiography}

\end{document}